\documentclass[sigconf]{acmart}
\AtBeginDocument{%
  }

\acmSubmissionID{8606}

\usepackage{enumitem}
\usepackage{booktabs}
\usepackage{multirow}
\usepackage{colortbl}
\usepackage{makecell}

\newcommand{\systemname}{AskEase}
\definecolor{lightyellow}{HTML}{FBEDC1}
\definecolor{lightgreen}{HTML}{D2ECC7}
\definecolor{lightpink}{HTML}{F6DFDF}
\definecolor{placeholder}{HTML}{BE3D27}

\newcommand{\nan}[1]{\textcolor{black}{#1}}
\newcommand{\fix}[1]{\textcolor{black}{#1}}
\newcommand{\final}[1]{\textcolor{black}{#1}}

\newcommand{\timeout}[1]{\textcolor{orange}{#1}}
\newcommand{\fail}[1]{\textcolor{orange}{#1}}
\newcommand{\gaveup}[1]{\textcolor{orange}{#1}}

\copyrightyear{2026}
\acmYear{2026}
\setcopyright{cc}
\setcctype{by}
\acmConference[CHI '26]{Proceedings of the 2026 CHI Conference on Human Factors in Computing Systems}{April 13--17, 2026}{Barcelona, Spain}
\acmBooktitle{Proceedings of the 2026 CHI Conference on Human Factors in Computing Systems (CHI '26), April 13--17, 2026, Barcelona, Spain}
\acmDOI{10.1145/3772318.3790661}
\acmISBN{979-8-4007-2278-3/2026/04}

\begin{document}

\title[Context-Aware Guidance for Screen Reader Users in Computer Use]{From Struggle to Success: Context-Aware Guidance for Screen Reader Users in Computer Use}



\author{Nan Chen}
\affiliation{%
  \institution{Microsoft Research}
  \city{Shanghai}
  \country{China}}
\email{nanchen@microsoft.com}

\author{Jing Lu}
\authornote{Work done during internship at Microsoft Research.}
\affiliation{%
  \institution{College of Computer Science and Artificial Intelligence, Fudan University}
  \city{Shanghai}
  \country{China}}
\email{jinglu24@m.fudan.edu.cn}

\author{Zilong Wang}
\authornote{corresponding author.}
\affiliation{%
  \institution{Microsoft Research}
  \city{Shanghai}
  \country{China}}
\email{wangzilong@microsoft.com}

\author{Luna K. Qiu}
\affiliation{%
  \institution{Microsoft Research}
  \city{Shanghai}
  \country{China}}
\email{lunaqiu@microsoft.com}

\author{Siming Chen}
\authornotemark[2]
\affiliation{%
  \institution{School of Data Science, Fudan University}
  \city{Shanghai}
  \country{China}}
\email{simingchen@fudan.edu.cn}

\author{Yuqing Yang}
\authornotemark[2]
\affiliation{%
  \institution{Microsoft Research}
  \city{Shanghai}
  \country{China}}
\email{yuqyang@microsoft.com}


\renewcommand{\sectionautorefname}{Section}
\renewcommand{\subsectionautorefname}{Section}
\renewcommand{\subsubsectionautorefname}{Section}

\begin{abstract}

Equal access to digital technologies is critical for education, employment, and social participation.
However, mainstream interfaces are visually oriented, creating steep learning curves and frequent obstacles for screen reader users, and limiting their independence and opportunities.
Existing support is inadequate---tutorials mainly target sighted users, while human assistance lacks real-time availability.
We introduce AskEase, an on-demand AI assistant that provides step-by-step, screen reader user-friendly guidance for computer use. 
AskEase manages multiple sources of context to infer user intent and deliver precise, situation-specific guidance.
Its seamless interaction design minimizes disruption and reduces the effort of seeking help.
We demonstrated its effectiveness through representative usage scenarios and robustness tests.
In a within-subjects study with 12 screen reader users, AskEase significantly improved task success while reducing perceived workload, including physical demand, effort, and frustration.
These results demonstrate the potential of LLM-powered assistants to promote accessible computing and expand opportunities for users with visual impairments. 


\end{abstract}

\begin{teaserfigure}
    \begin{center}
    \includegraphics[width=0.95\textwidth]{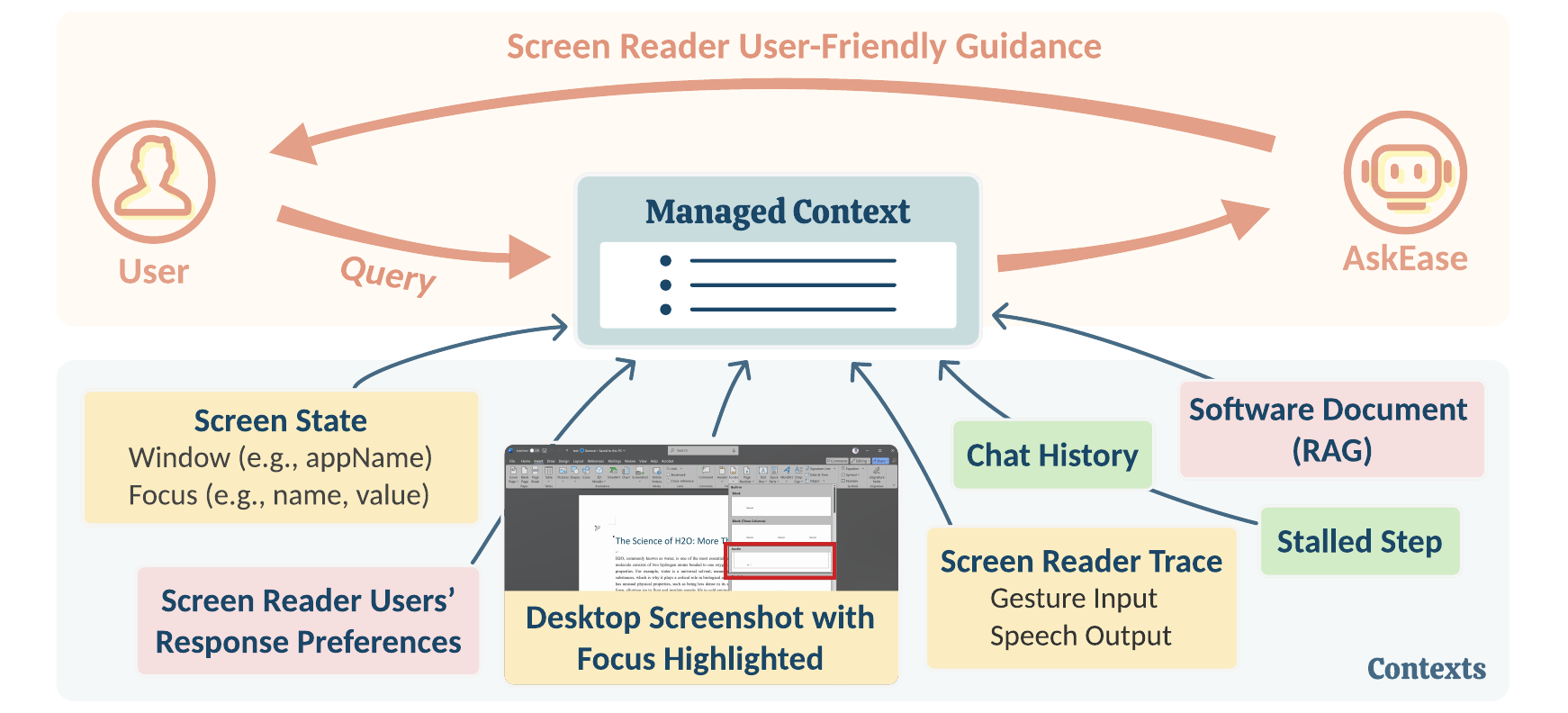}
    \end{center}
    \vspace{-1em}
  \caption{
      \systemname{} automatically collects and manages multiple sources of context to enhance situational awareness and provide precise, screen reader user-friendly guidance. It considers three main types of context: 
      (1) \raisebox{0pt}[0pt][0pt]{\colorbox{lightyellow}{Environment contexts}}, including desktop screenshots with focused elements highlighted, structured screen states, and live screen reader traces; 
      (2) \raisebox{0pt}[0pt][0pt]{\colorbox{lightpink}{Knowledge contexts}}, such as retrieved software documentation and response preference principles for screen reader users; 
      (3) \raisebox{0pt}[0pt][0pt]{\colorbox{lightgreen}{Conversational contexts}}, comprising chat history and the stalled step. 
   }
  \Description{Architecture of AskEase. At the top, curved arrows show the interaction flow: a User sends a "Query" through a central "Managed Context" hub to AskEase, which then provides "Screen Reader User-Friendly Guidance" to the user. The central "Managed Context" hub manages inputs from seven context sources. These sources include: "Screen State" (containing window and focus information), "Desktop Screenshot with Focus Highlighted" (an actual screenshot with a highlighted focus element), "Screen Reader Trace" (listing gesture input and speech output), "Chat History", "Stalled Step", "Screen Reader Users' Response Preferences", and "Software Document (RAG)".}
  \label{fig:teaser}
\end{teaserfigure}


\maketitle

\section{Introduction}
\label{sec-intro}

In contemporary society, access to computers and digital services underpins opportunities in education, employment, and social participation~\cite{puryear2022github,adeshola2024opportunities,atcheson2025d, wahidin2018challenges},
yet achieving universal access remains a persistent challenge for inclusive interactive systems~\cite{billah2017ubiquitous,vanderheiden2008ubiquitous}. 
Users with visual impairments rely primarily on screen readers (SRs) to convert digital content into speech and enable keyboard-based navigation.
Despite advances in assistive technologies, SR users continue to face steep learning curves and frequent obstacles in computer use~\cite{wahidin2018challenges,storer2021s,mcdonnall2024actual,perera2025sky}.
Mainstream interfaces are visually oriented, forcing SR users to memorize multiple keyboard shortcuts and prior screen reader output.
These constraints impose substantial cognitive load and slow down essential activities like navigation~\cite{chheda2025brought,rodrigues2020open,saha2023understanding}, element discovery~\cite{potluri2018codetalk,rodrigues2017context}, overview formation~\cite{mahmud2007combating,ramakrishnan2017non}, and change detection~\cite{potluri2018codetalk}.
These difficulties persist even with emerging AI assistants~\cite{adnin2024look}. 
While tools such as coding assistants~\cite{chen2025screenreaderusersvibe,flores2025impact}, facilitate code generation, they also pose challenges for content tracking, given their dynamic responses and the need for frequent window switching.

When encountering difficulties with digital systems, SR users often draw on online tutorials~\cite{mcdonnall2024actual,perera2025sky} or peer assistance~\cite{saha2023understanding,vigo2013coping,perera2025sky,rodrigues2017context} to overcome obstacles.
Prior HCI research has supported these strategies by facilitating accessible tutorial creation~\cite{saha2023tutoria11y,rodrigues2021promoting} and streamlining help-seeking~\cite{rodrigues2017context}.
For example, Tutoria11y~\cite{saha2023tutoria11y} helps creators produce step-by-step audio guidance by streamlining narration and screen reader speech recording, while Rodrigues et al.~\cite{rodrigues2017context} introduced a human-powered Q\&A service leveraging screenshots and interface details to support user queries.
However, these approaches ultimately rely on human contributors with domain expertise in SR interaction to provide answers and guidance~\cite{rodrigues2017context,saha2023tutoria11y}, which constrains their scalability and responsiveness.

Recent advances in Large Language Models (LLMs) offer promising avenues to address these challenges by providing real-time learning support~\cite{chatgpt_study_mode,laato2023ai} and troubleshooting assistance~\cite{marek2024chatgpt,khurana2024and}.
Although LLM-based assistants with step-by-step visual guidance have proven effective for learning and skill development ~\cite{khurana2025me,khurana2024and}, they fail to address SR users' specific needs. 
Current AI assistants pose a dual challenge for SR users. First, their guidance is typically expressed using visual descriptions and mouse-based interactions. Second, they require users to formulate prompts and supply contextual information to enable the AI to accurately interpret user intentions~\cite{subramonyam2024bridging,bansal2024challenges,chen2025screenreaderusersvibe,zamfirescu2023why}. 
These factors impose additional cognitive and operational burdens on SR users~\cite{chen2025screenreaderusersvibe,perera2025sky,adnin2024look}, highlighting the need for on-demand assistants that provide precise and accessible guidance with minimal effort.

We present \systemname{}, a real-time help-seeking assistant that supports computer use for SR users.
For example, when they encounter confusion or uncertainty, \systemname{} can provide step-by-step guidance to help users discover hidden elements, interpret complex or inaccessible screen content, and approach frustrating tasks independently and confidently.
\systemname{} automatically collects and manages rich contexts---screen states, desktop screenshots, live screen reader traces, and recent chat history (see \autoref{fig:teaser})---to analyze the current interface, understand user challenges, and infer user intent.
It further leverages external documentation via retrieval-augmented generation (RAG) and applies screen reader-specific response principles to deliver accurate, accessible, and easy-to-follow guidance.
Its seamless interaction design further minimizes the effort required to get help---users can request and review adaptive guidance without disrupting their workflow.

We prototyped \systemname{}\footnote{\final{The source code is available at \url{https://github.com/microsoft/AskEase}.}} as an LLM-powered NVDA~\cite{nvda} add-on. Its effectiveness was demonstrated through two representative usage scenarios---support for learning emerging AI tools and assistance in addressing accessibility issues---together with robustness tests across 12 applications.
We further conducted a within-subject study with 12 SR users, where participants completed unfamiliar tasks using either \systemname{} or their usual tools (e.g., search engines, AI assistants).
Results indicate that \systemname{} improved task completion rates while reducing perceived workload, including physical demand, effort, and frustration.
Participants highlighted that \systemname{}'s accessible design enabled efficient help-seeking and reduced obstacles to engaging with computer systems.

Overall, our work investigates how cutting-edge technologies can bridge the gap to ubiquitous accessibility, enabling people with visual impairments to interact with computing systems with greater confidence and expand their societal and professional opportunities. Our contributions are threefold:
\begin{itemize}[left=0pt]
    \item We identify five design goals for assistive systems supporting computer use by SR users through a participatory formative study.
    \item We introduce \systemname{}, an on-demand assistant that automatically collects and manages multiple sources of context to enhance situational awareness and provide precise, screen reader user-friendly guidance. Its effectiveness is demonstrated through two realistic user scenarios and robustness testing across 12 applications.
    \item We conduct a within-subject study with 12 SR users, showing how \systemname{} supports engagement with computing tasks, and discuss key observations that inform future approaches to accessible computing.
\end{itemize}


\section{Related Work}
\label{sec-related}
\subsection{Challenges in Computer Use for SR users}

For SR users, computer use involves layered challenges, ranging from mastering screen reader operation and its complex navigation commands~\cite{mcdonnall2024actual,wahidin2018challenges} to coping with poorly structured interfaces and insufficient accessibility design that hinder accurate access to digital content~\cite{das2019doesn,momotaz2021understanding,lazar2007frustrates,vigo2013coping,rodrigues2017context,potluri2018codetalk,rodrigues2020open}.
Challenges are further complicated when learning and using feature-rich applications~\cite{hamideh2024accessibility,moore2018keyboard,schaadhardt2021understanding}---such as word processors~\cite{perera2025sky,momotaz2021understanding,buzzi2010accessing}, programming environments~\cite{chen2025screenreaderusersvibe,potluri2018codetalk}, and audio production tools~\cite{saha2023understanding}.
Users may struggle to explore all available functionalities due to the sheer number of elements~\cite{potluri2018codetalk,rodrigues2017context}.
Moreover, the sequential nature of auditory output enforces a linear information flow that limits their ability to develop an overall understanding of the application~\cite{mahmud2007combating,ramakrishnan2017non}. 
These factors often lead to navigation challenges, including disorientation when switching panels and inefficiency in locating feature elements~\cite{potluri2018codetalk,chheda2025brought,rodrigues2020open,saha2023understanding}.
Consequently, users face uncertainty about whether task failures arise from inaccessible information, access friction, or true absence of the needed content~\cite{bigham2017effects}.
The need to memorize extensive shortcuts for efficient operation further increases cognitive load~\cite{ramakrishnan2017non,kodandaram2024enabling}.
Accessibility challenges remain in emerging AI assistants.
For example, users have difficulty tracking AI generation status and become disoriented when reviewing outputs across multiple windows~\cite{chen2025screenreaderusersvibe,flores2025impact,adnin2024look}.
These persistent barriers highlight the need for intelligent systems that empower SR users to navigate interfaces confidently, learn applications effectively, and resolve difficulties with minimal friction.

\subsection{Accessible Learning Support and Help-Seeking for Screen Reader Users}

Learning support resources, including tutorials and guidance materials, play a crucial role in computer use and software learning~\cite{chilana2012lemonaid,grossman2009survey,khurana2024and,novick2006don,kim2014crowdsourcing}.
However, these resources are often inaccessible to SR users. 
Most learning resources rely heavily on visual cues and rarely provide screen reader or keyboard-oriented instructions~\cite{chen2025screenreaderusersvibe,perera2025sky,storer2021s}. 
To improve accessibility, Rodrigues et al.~\cite{rodrigues2019understanding} recommend avoiding visual-only references and including textual labels and interface element purposes.
Systems like Tutoria11y~\cite{saha2023tutoria11y} and RISA~\cite{rodrigues2021promoting} attempt to lower tutorial creation barriers, with Tutoria11y enabling audio tutorials through \nan{captured voice instructions and automatic detection of underlying application state changes}.
Despite these advances, accessible tutorial availability remains sparse due to manual authoring constraints.

Given limited accessible resources, SR users frequently seek help through online communities.
Their help-seeking behaviors exhibit distinctive characteristics: they provide detailed task descriptions and screen reader activity traces to build shared understanding~\cite{saha2023understanding}, and frame questions with records of attempted steps~\cite{johnson2022program}.
Posts in accessibility forums use shorter sentences and higher readability, reflecting deliberate adaptation for better comprehension~\cite{venkatraman2024you}.
However, such practices can be burdensome, requiring extensive contextual information, and sighted peers often lack screen reader expertise~\cite{rodrigues2017context,saha2023tutoria11y}.
To address context-sharing challenges, HintMe!~\cite{rodrigues2017context} augments user queries with screenshots and interface details for volunteer assistance.

\nan{Although earlier approaches were constrained in scalability and failed to offer real-time support, previous studies have provided guidance on delivering SR user-friendly responses~\cite{saha2023tutoria11y,rodrigues2017context,saha2023understanding,rodrigues2019understanding,venkatraman2024you} and have highlighted the value of rich contextual information for interpreting user intent and situational challenges~\cite{rodrigues2017context,saha2023understanding,johnson2022program}.}
Motivated by these findings, we propose an on-demand assistant that automatically gathers multiple sources of context to enhance situational awareness and provides accessible support aligned with SR user practices.

\subsection{LLM-Assisted Computer Learning and Use}

The emergence of LLMs has led to numerous AI assistants~\cite{sellen2024rise,afzoon2025modeling}, such as Microsoft 365 Copilot~\cite{microsoft365}, ChatGPT~\cite{openai_chatgpt}, and Claude~\cite{anthropic_claude}. These assistants provide computer support through two primary paradigms: task automation, where assistants directly execute user requests through natural language, proving particularly effective for simple, repetitive tasks~\cite{khurana2025me,flores2025impact}; and instructional guidance, which provides step-by-step explanations and scaffolded support, empowering users to accomplish tasks independently while building their competence~\cite{khurana2025me,khurana2024and}.
By providing processes rather than just outcomes, the instructional guidance enables users to develop understanding through hands-on practice~\cite{carroll1987paradox,kiani2019beyond,khurana2024and}, helps build robust mental models essential for effective software control~\cite{bansal2024challenges,nourani2021anchoring}, reduces the risk of overreliance on AI systems~\cite{qiao2025use,spatharioti2025effects,bo2025s}, and develops domain knowledge that enables users to better evaluate AI responses~\cite{khurana2024and,shneiderman1997direct}.
%
Studies have demonstrated that well-designed step-by-step visual guidance can significantly improve both user control and learnability, particularly in complex, feature-rich applications~\cite{khurana2025me}.
However, these existing approaches\nan{~\cite{khurana2025me,khurana2024and,li2025explorar,yang2024aqua} mainly} target sighted users, relying on visual cues that are inaccessible to SR users. 

While some recent efforts have begun addressing accessibility concerns with LLMs~\cite{kodandaram2024enabling,xie2025beyond}---providing personalized visual guidance for low-vision learners~\cite{sechayk2025veasyguide}, enhancing video game accessibility~\cite{qiu2025gamerastra}, and improving task-specific web navigation~\cite{mohanbabu2025task}---\nan{other work such as Vid2Coach~\cite{huh2025vid2coach} translates how-to videos into narrated instructions tailored for blind and low-vision users in cooking tasks using RAG. However, these approaches are tied to specific domain contexts or media (e.g., video tutorials, smart glasses) and do not generalize to on-demand support for SR users in general computer use.}
To address this gap, we conducted a formative study to identify design goals and implemented an assistant tailored for SR users, \nan{providing step-by-step instructional guidance aligned with their interaction practices.}
%
%
%


\subsection{\fix{AI-based Accessibility Overlays}}

\nan{Accessibility overlays have been developed to automatically repair accessibility barriers in websites and applications by dynamically modifying the rendered interface, such as adding alternative text, adjusting color contrast, or enabling keyboard navigation~\cite{overlayfactsheet, nacheva2023heuristic, egger2022overlay}.
While these tools can provide immediate, surface-level improvements, empirical studies show that they are often unreliable: overlays may conflict with assistive technologies such as screen readers~\cite{makati2024Promise}, mask persistent accessibility barriers~\cite{hartman2025evaluating}, and create a false sense of compliance~\cite{EDF2023_overlays,overlayfactsheet}.
Moreover, reliance on overlays has been criticized for potentially reducing companies' responsibility for maintaining long-term accessibility~\cite{makati2024Promise, overlayfactsheet}.}

\nan{
In contrast, \systemname{} adopts a user-centered approach to meet the needs of SR users. 
\systemname{} follows SR conventions and aligns with users' SR workflows~\cite{perera2025sky,phutane2023speaking}.
It provides non-intrusive, step-by-step guidance that preserves the original interface, facilitating exploration and interaction with both standard and well-accessible applications.
The assistant leverages rich contexts to provide robust and contextually relevant guidance, enabling users to benefit from existing accessibility features and enhancing independent software use without masking underlying accessibility barriers.
}

\section{\systemname}
\label{sec-design}
To better understand how to support SR users in computer use, 
we developed an initial prototype informed by prior research on computer use and help-seeking challenges. 
We then conducted a participatory formative study with three SR users to gather feedback on functionality, uncover unmet needs, and refine the interaction design. 
The insights guided the design goals presented below.

\subsection{Initial Prototype}
Screen readers (SRs) enable visually impaired users to access digital interfaces by converting visual content into auditory output. 
In daily usage scenarios, SR users, who are blind or low-vision, navigate complex interfaces and operate features through numerous keyboard commands, needing to remember shortcuts, previous steps, and element positions, since they cannot rely on visual cues to guide their actions~\cite{bigham2008webanywhere}.
This process can be demanding with high cognitive loads: complex interfaces require listening to long streams of information to locate target elements, accessibility issues may prevent identification of some items, dynamic updates can go unnoticed, and accidental focus shifts may leave users unsure how to recover~\cite{lazar2007frustrates}. 
Prior research has shown that these interactions can lead to disorientation, uncertainty about next steps, and missed interface changes~\cite{rodrigues2017context, potluri2018codetalk, ramakrishnan2017non, bigham2017effects, chen2025screenreaderusersvibe, flores2025impact}.
These challenges underscore the barriers SR users face in daily computer use and motivate the design of accessible assistants that enable equal access.

Motivated by these observations, we designed the initial prototype of the computer-use assistant, which provides real-time descriptions of interface and suggests possible actions to help users overcome challenges and lower barriers to learning new applications.
It was implemented as an NVDA~\cite{nvda} add-on, taking advantage of NVDA’s widespread adoption among blind and visually impaired users and its open-source architecture for rapid development~\cite{momotaz2023understanding}.
The prototype offers two core features: (1) \emph{Question \& Answer} enables an interactive dialogue in which users can pose free-form questions about computer use, such as discovering unfamiliar functionality or obtaining procedural guidance. The assistant responds in HTML format, enabling the output to be rendered with structured headings and contextual information for efficient navigation~\cite{Borodin2010More,jordan2024information} (2) \emph{Screen Description} allows users to press a single shortcut key to obtain a structured summary of the current computing context, such as the active window and focus, to support rapid reorientation, a need frequently reported in prior work~\cite{vigo2013coping,lunn2011identifying,chen2025screenreaderusersvibe}.

\subsection{Participatory Formative Study}
We conducted a remote participatory formative study with three experienced screen reader (SR) users recruited through research networks: an open-source screen reader contributor (C1), an accessibility optimization specialist (C2), and a user with extensive experience and interest in AI (C3). 
All participants were blind, and their diverse backgrounds provided complementary perspectives spanning infrastructure development, applied accessibility practice, and forward-looking engagement with emerging AI technologies.

Each session lasted approximately one hour and consisted of three parts: (1) a brief introduction to the prototype; (2) free-form exploration during which participants performed self-selected routine tasks; and (3) a semi-structured interview to gather feedback on their experiences, including perceived utility, encountered limitations, desired refinements, and potential opportunities. 
Structured notes were taken to capture participants' confusion points, feature requests, and expectations regarding context use and response style. 
Each participant received \$30 USD (or local equivalent) as compensation for their time and effort. The study protocol was approved by our Institutional Review Board (IRB).

\subsection{Design Goals}
\label{sec:design-goal}
All participants highlighted \systemname{}'s potential to reduce adoption barriers and enable more deliberate interaction. 
Building on these findings and prior work on SR users' help-seeking behaviors, we present the following design goals:

\begin{itemize}[left=0pt]
\item \textbf{DG1. Situation Awareness.} 
Support understanding of the user's current situation and intent~\cite{bansal2024challenges,khurana2024and}, providing a contextual foundation for precise assistance (C1, C2, C3). This involves capturing relevant screen information and software state to inform the assistant's reasoning.

\item \textbf{DG2. Preference-Aligned Responses.} 
Ensure that guidance aligns with users' expectations for accessible interaction~\cite{rodrigues2017context,saha2023tutoria11y,rodrigues2019understanding,saha2023understanding,venkatraman2024you}, such as using keyboard-friendly actions and describing elements without relying on visual cues (C1, C3). This involves understanding and adapting to user-specific preferences.

\item \textbf{DG3. Actionable and Correct Guidance.} 
Provide step-by-step guidance that is feasible, accurate, and directly supports the user overcoming interaction difficulties (C2, C3). 
Participants noted that while contextual descriptions were useful, guidance that actually resolves the difficulty was considered more valuable. Relevant documents can be integrated to generate such guidance.

\item \textbf{DG4. Seamless In-Flow Assistance.} 
Provide help that preserves task flow and attentional focus by minimizing focus shifts and window switching (C2, C3). 
Participants reported that frequent switching to external help windows disrupted workflow and increased refocusing effort~\cite{chen2025screenreaderusersvibe, flores2025impact}.

\item \textbf{DG5. Accessible, Fine-Grained Interaction Design.} 
Enhance efficiency and reduce cognitive load through fine-grained interaction design, ensuring accessibility (C1, C2). 
Participants suggested several specific strategies: presenting one actionable unit per list item to support both character-by-character review and rapid line skimming, grouping or collapsing relevant elements to shorten Tab traversal, and using familiar shortcuts (e.g., Escape to dismiss dialogs) to reduce onboarding effort.

\end{itemize}

\section{System}
\label{sec-method}
Building on our design goals (\autoref{sec:design-goal}), we developed \systemname{}, an intelligent on-demand assistant that empowers SR users in computer use. 
Its interface and interaction design (\autoref{sec:interface}) delivers low-effort, contextually relevant assistance while minimizing workflow disruption. 
The context engineering (\autoref{sec:context-engineering}) captures and manages rich, task-relevant information to ground the assistant's understanding of user intent and interface state, enabling accurate, user-aligned guidance. 
We illustrate \systemname{}'s functionality through two representative usage scenarios (\autoref{sec:user-scenario}), showing how it supports efficient interaction across modern AI tools and productivity tasks. Finally, we evaluate its robustness and efficiency across 12 applications (\autoref{sec:robustness-evaluation}), demonstrating its effectiveness in assisting SR users.

\subsection{Interface and Interaction Design}
\label{sec:interface}

In this section, we present the interface and interaction design of \systemname{}, showing how its on-demand \emph{contextual Q\&A}, \emph{adaptive support}, and \emph{screen description} features deliver assistance with minimal workflow disruption for SR users.

\subsubsection{Contextual Question \& Answer.}
\begin{figure}[htbp]
    \centering
    \includegraphics[width=0.95\linewidth]{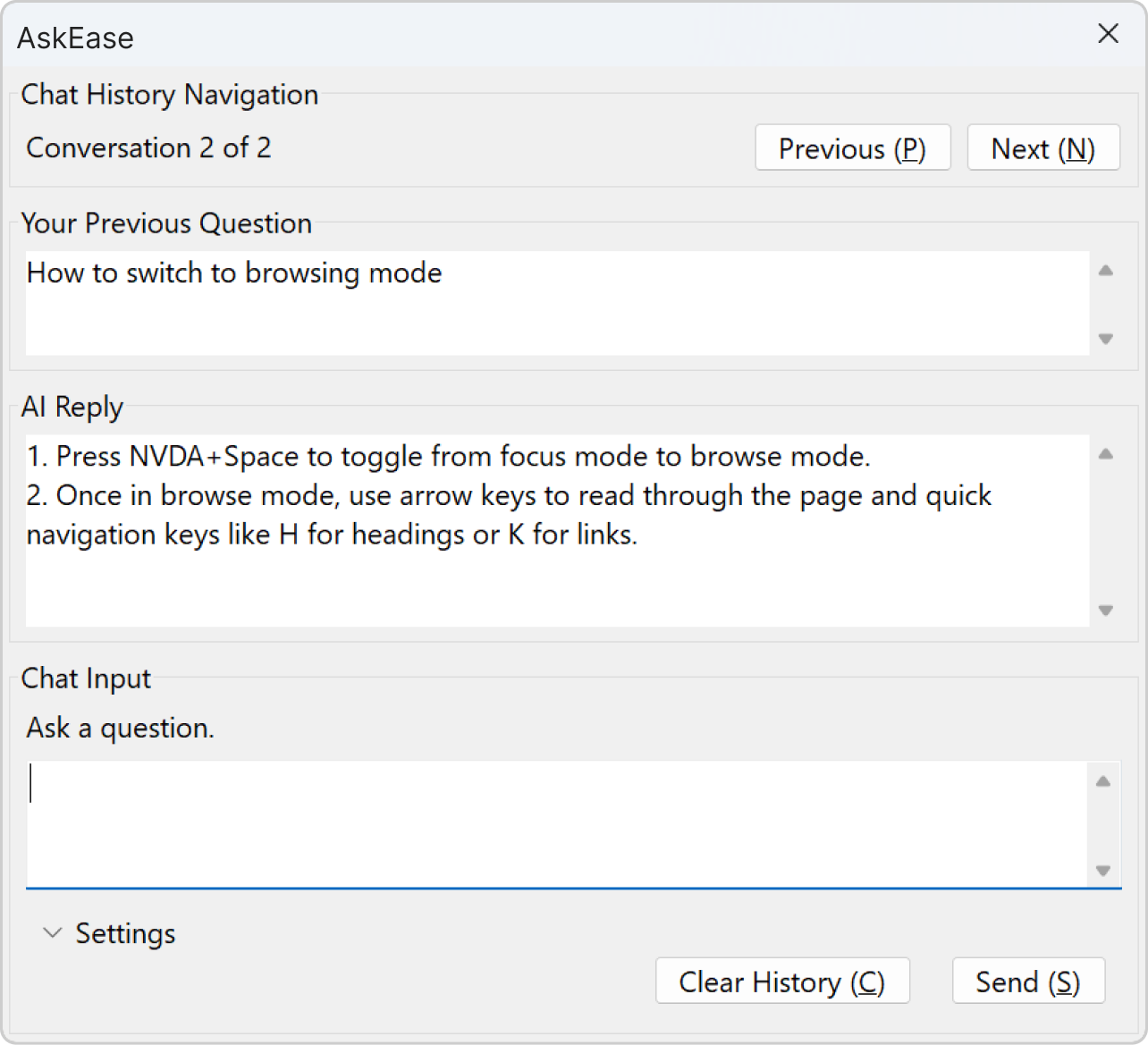}
    \caption{\systemname{} dialog: pop-up with a question input field and chat history view\nan{, showing one user question and its AI reply at a time, with Previous/Next buttons to navigate through the multi‑turn chat history.}.}
    \Description{AskEase dialog interface with several main sections: "Chat History Navigation" at top showing "Conversation 2 of 2" with "Previous (P)" and "Next (N)" buttons. The main area displays a previous interaction - user question "How to switch to browsing mode" and AI reply with two steps for using NVDA+Space and navigation keys. Bottom sections include a "Chat Input" area with an empty text field, collapsed "Settings", "Clear History (C)", and "Send (S)" buttons.}
    \label{fig:interface}
\end{figure}

When users feel confusion or frustration, they can invoke the \emph{Contextual Question \& Answer (Contextual Q\&A)} feature, which opens a streamlined dialog (\autoref{fig:interface}) for entering a question and reviewing the response.  
The assistant automatically captures and manages relevant context (see \autoref{sec:context-engineering}) to analyze the situation and infer user intent (\textbf{DG1}). 
It also retrieves relevant documentation to generate more accurate answers (\textbf{DG2}).
During response generation, the assistant provides textual cues and audio cues to communicate AI status, and automatically announces the generated response to users (\textbf{DG5}).  
Responses are presented as a numbered list of clear, keyboard-focused steps, avoiding visual-only descriptions (\textbf{DG2}).  
The dialog is hidden automatically by default, restoring prior focus.  
Users can navigate \fix{the generated step‑by‑step guidance} using dedicated \emph{Previous Step} and \emph{Next Step} \fix{keyboard} shortcuts, reviewing \fix{each step} at their own pace without losing focus on the primary task (\textbf{DG4}).

\subsubsection{Built-in Support Features}
The \systemname{} offers two built-in features that lower the effort of seeking help.

\begin{itemize}[left=5pt]
    \item \emph{Adaptive Support.} When users struggle to follow provided steps, they can trigger this feature to receive intelligently adaptive guidance (\textbf{DG3}). The assistant will analyze the managed context (e.g., SR trace, screen state, detailed in~\autoref{sec:context-engineering}) to identify user's confusion, then generate more feasible, context-aware guidance (\textbf{DG1}).

    \item \emph{Screen Description.} When users feel disoriented, they can invoke this feature to receive a concise, structured summary of their active windows and current focus and relevant suggestions to support rapid reorientation.

\end{itemize}


\subsubsection{Seamless Help-Seeking Flow}
Together, the assistant's features support accessible, low-effort help-seeking (\textbf{DG4}) by preserving task flow and minimizing disruption. 
They achieve this through three key strategies: automatically maintaining contextual awareness to infer user intent and provide targeted assistance, ensuring focus continuity so users can navigate guidance step by step without losing track of their work, and enabling in-place invocation of built-in features via shortcuts, so users can quickly access support without entering a question.

\subsection{Context Engineering}
\label{sec:context-engineering}


Recent advances in Multimodal Large Language Models (MLLMs) have demonstrated remarkable capabilities in graphical user interface (GUI) understanding~\cite{tang2025survey,zhang2024large}, reasoning~\cite{plaat2024reasoning,cheng2025empowering}, and planning~\cite{huang2024understanding,wei2025plangenllms}, highlighting their suitability for supporting SR users. 
In interactions with MLLMs, \textbf{context} is crucial for bridging the gap between user intent and model comprehension~\cite{mei2025survey,gao2024taxonomy}, encompassing information about the user, their environment, and relevant knowledge. 
Prior work on help-seeking behaviors among SR users shows that they often provide rich contextual information to help others understand their situation~\cite{rodrigues2017context,saha2023understanding,johnson2022program}, suggesting a key design opportunity: automatically capturing and managing such information can reduce user effort while enabling more accurate assistance. 
Based on a systematic analysis of help-seeking patterns and LLM capabilities, we categorize context into three types---\emph{environment context}, \emph{knowledge context}, and \emph{conversational context} (\autoref{fig:teaser}).



\subsubsection{Environment Context}
Environmental context captures the current state of the user's computing environment, providing the assistant with situational awareness. We collect data from the screen and screen reader at the moment of each user query. This includes three key types:

\underline{\textit{Desktop Screenshot with Focus Highlighted.}}
Desktop screenshots provide a visual overview of the current screen~\cite{rodrigues2017context}, while highlighting the focused element indicates the user's current focus. 
Leveraging MLLMs' GUI understanding, these inputs enable the model to infer user intent, reason about interface layouts, and compensate for missing accessibility information, especially in poorly supported applications.
We use Python's Pillow library to capture screenshots and NVDA's API to retrieve focus coordinates, with a red border marking the focused element.

\underline{\textit{Screen State.}}
Screen state information, including active window and focused element properties, complements visual screenshots with semantic attributes from accessibility APIs. 
Combining these attributes with visual information maximizes information completeness.
For example, different versions of software may use different shortcuts; such information allows the assistant to provide more targeted responses.
Our implementation uses NVDA's API to extract active window properties (e.g., application name, version) and focused element details (e.g., type, name, role, keyboard shortcuts, help text, descriptions), formatting this as structured text.

\underline{\textit{Screen Reader Trace.}}
Prior research has shown the value of SR activity traces and user-described operational steps in supporting shared awareness between users and assistants~\cite{saha2023understanding,johnson2022program}. 
We find this information also helpful for enabling models to understand user operation sequences and diagnose issues.
For instance, when users struggle, the model can analyze which keystrokes were pressed and what speech output was heard to determine whether users pressed incorrect shortcuts, pressed correct shortcuts that failed to execute, or encountered discrepancies between generated guidance and actual screen reader speech that prevented successful completion.
We hook into NVDA's event and speech pipelines to capture gesture inputs and speech outputs, storing them in a chronological list.

\subsubsection{Knowledge Context}
\label{sec:knowledge-context}
Knowledge context---curated from human expertise and dynamically retrieved for relevance---injects external knowledge into the model, enabling more precise and user-aligned responses.

\underline{\textit{Screen Reader Users' Response Preferences.}}
Drawing from prior research on SR users' tutorial design and communication patterns in online forums, we established a set of guiding principles to ensure generated responses align with common SR user preferences:
\textit{(1)} Responses should prioritize keyboard-based interactions, the primary mode of operation for SR users~\cite{saha2023tutoria11y,rodrigues2017context,saha2023understanding}. 
\textit{(2)} Responses should avoid purely visual descriptions and instead reference specific GUI element names and types (as perceived by SR), along with navigational cues to support element location~\cite{saha2023tutoria11y,saha2023understanding,rodrigues2019understanding}.
\textit{(3)} Responses should be concise and readable. Prior work has shown that SR users prefer succinct language with fewer words per sentence due to the linear nature of audio consumption~\cite{venkatraman2024you}.
\textit{(4)} Responses should use standard terminology and avoid ambiguous vocabulary, helping users build accurate mental models of system functionality~\cite{saha2023understanding}.
These principles are included through prompt engineering~\cite{schulhoff2024prompt,marvin2023prompt} to guide model responses according to users' preferences.

\underline{\textit{Software Documentation.}}
Although MLLMs contain some embedded knowledge of software usage from pretraining, such knowledge can be stale, partial, or incomplete. 
To provide up-to-date information, we develop a lightweight RAG framework that integrates software documentation into a structured knowledge base, supporting efficient semantic retrieval.
We organize the RAG framework construction into three main steps, detailed below.

\textit{Chunking:} We manually collected official documentation (e.g., help documents and FAQs), prioritizing materials for SR users with non-visual descriptions and keyboard-based instructions; otherwise, general documentation was used to support model understanding. The dataset covers 14 applications used in robustness evaluation (\autoref{sec:robustness-evaluation}) and user study (\autoref{sec:user-study-app-task}). 
All text content in the documentation is segmented into coherent chunks based on subheadings, preserving the software name, section titles, and content.

\textit{Indexing:}
To support queries expressed in different wordings or languages, \texttt{GPT-5-mini} generates five \final{paraphrased} variants per chunk for each supported language (e.g., English and Chinese)\final{, capturing lexical and syntactic variations that users might employ when referring to the same content.}
Each chunk and its variants are encoded into embeddings using \texttt{text-embedding-3-large} and linked to the same chunk ID.
The embeddings are stored with metadata in a SQLite database and indexed in FAISS (\texttt{IndexFlatIP}) for efficient similarity search.


\textit{Querying:} 
User queries are embedded and searched against the FAISS index.
\final{\systemname{} optionally supports query expansion via HyDE~\cite{gao2023precise} to improve retrieval recall. This expansion is disabled by default to reduce latency and overhead, and is intended for cases where initial retrieval is insufficient or when users trade response time for higher accuracy.}
Embeddings from both the original and expanded queries are independently used for retrieval. Results corresponding to the same chunk ID are merged and scored via max pooling to produce a final ranked list. The most relevant chunks are then used as context for the assistant to generate guidance.

\subsubsection{Conversational Context}
Conversational context refers to information from ongoing user-assistant interactions that helps maintain continuity and coherence across multiple turns.

\underline{\textit{Chat History.}}
We retain chat history, including user queries, context, and assistant responses. It enables the assistant to understand users' evolving needs and provide coherent guidance across turns.

\underline{\textit{Stalled Step.}}
\nan{
We define the stalled step as the step the user is currently viewing when \emph{Adaptive Support} is invoked, i.e., the most recent step reached via the \emph{Previous Step} or \emph{Next Step} shortcut. When users request assistance on a step, we infer that they may be stuck on that step and treat this navigation position as an indicator of a task breakdown.
}

\subsubsection{Context Engineering Implementation}
The implementation employs a structured prompt framework, in which contextual elements are treated as variables within carefully designed templates. 
Each template guides the model through systematic processing, from assessing the current state to integrating relevant knowledge and generating responses. 
These context combinations were refined through iterative empirical testing during system development. 
SR users' response preferences are encoded within the system instruction to ensure that all outputs adhere to accessibility principles. 
The \emph{Contextual Q\&A} feature combines desktop screenshots, screen state, retrieved software documentation, and chat history to generate reliable guidance. Additionally, the \emph{Adaptive Support} feature incorporates desktop screenshots, screen state, SR traces, chat history, and stalled step information to analyze user difficulties and provide adaptive guidance.
\nan{To further improve reliability, we also incorporate concise uncertainty-handling rules into the system instruction: the model should explicitly state uncertainty rather than guess, and request clarification when a query is ambiguous.}
To address token limitations, the system applies threshold-based truncation to lengthy contextual information. 
The system instruction and prompt templates are provided in the supplementary materials.

\subsection{User Scenario}
\label{sec:user-scenario}
To illustrate how our tool supports real-world computer use, we present two representative scenarios: one demonstrating how it helps users \textbf{learn emerging AI tools}, and another showing how it assists in \textbf{addressing accessibility issues}.
\subsubsection{GitHub Copilot---Agent Mode}

\begin{figure*}[htbp]
    \centering
    \includegraphics[width=\linewidth]{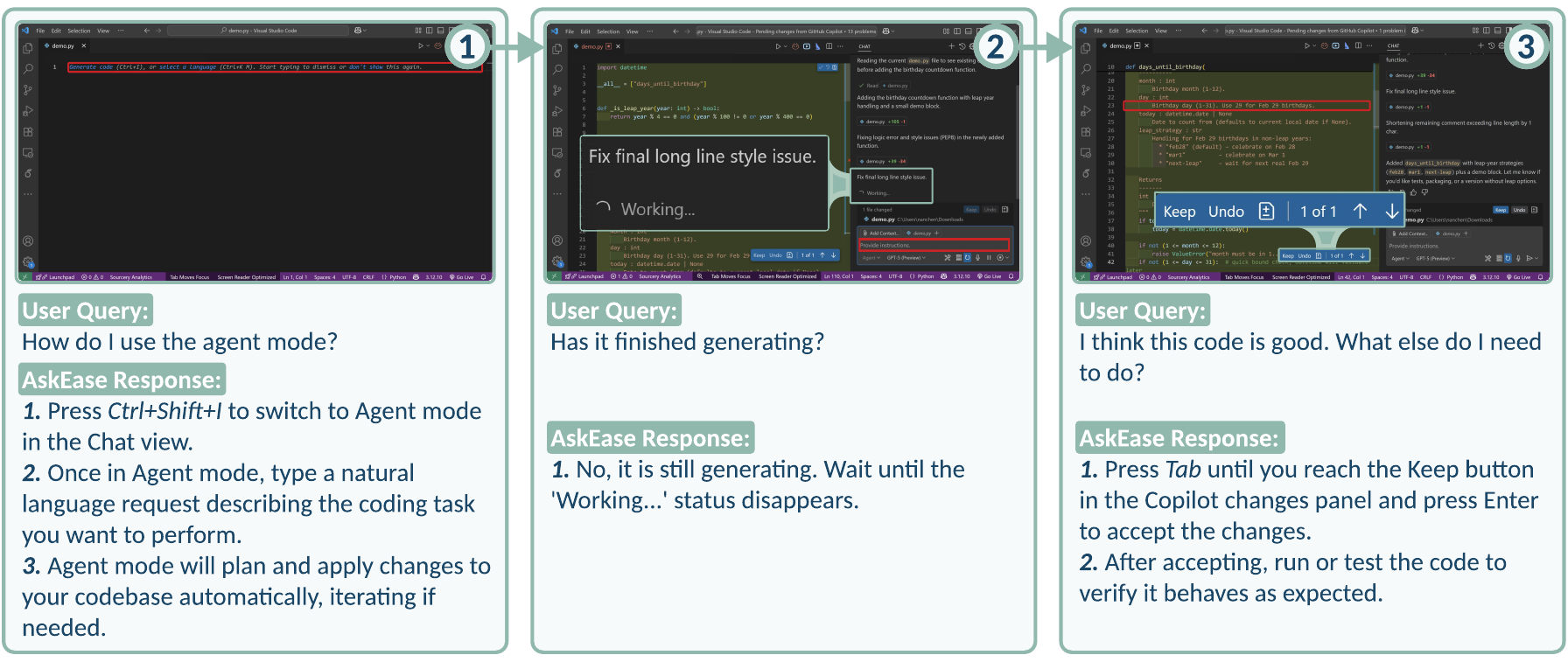}
    \caption{This scenario illustrates how \systemname{} supports users in learning and interacting with GitHub Copilot, an \textbf{emerging AI tool}. The interaction unfolds across three turns: onboarding, progress clarification, and actionable follow-up. }
    \Description{Three-panel sequence: Panel 1 displays an empty editor with a user query "How do I use the agent mode?" and AskEase response providing three-step instructions: press Control+Shift+I to activate Agent mode in Chat view, type a natural language request describing coding task, and agent mode will plan and apply changes automatically with iteration if needed. Panel 2 shows active code generation with "Working" status dialog, user asking "Has it finished generating?", and AskEase confirming generation is still in progress and to wait until the Working status disappears. Panel 3 presents generated code with Copilot changes panel showing "Keep" and "Undo" buttons, user stating "I think this code is good. What else do I need to do?", and AskEase providing two-step guidance: press Tab until reaching the Keep button and press Enter to accept changes, then run or test code to verify expected behavior.}
    \label{fig:copilot}
\end{figure*}

Mike is a blind developer who learns about the new \emph{agent mode} in \emph{GitHub Copilot}, which can assist in generating code automatically.
Unsure how to use this feature, he opens VS Code and invokes \emph{Contextual Q\&A}, asking ``How do I use the agent mode?'' (\autoref{fig:copilot} (1)). The assistant automatically organizes the contexts and provides step-by-step guidance. 
Following these instructions, Mike enables the agent mode and sends his request to Copilot.
After several minutes, Copilot continues to announce ``Working'', leaving Mike uncertain about the system's status. 
He then invokes \emph{Contextual Q\&A} again to seek clarification (\autoref{fig:copilot} (2)). 
The assistant confirms that code generation is still in progress, which reassures Mike and increases his confidence in understanding the situation.
Once Copilot completes the code generation, Mike reviews the suggested code and finds it satisfactory. He then queries \emph{Contextual Q\&A} about the next steps (\autoref{fig:copilot} (3)). The assistant guides him to accept and run the code. 
By following the assistant's instructions, Mike becomes familiar with the new agent mode and explores its features more confidently.

\subsubsection{Microsoft Word---Add Page Number}
\begin{figure*}[htbp]
    \centering
    \includegraphics[width=\linewidth]{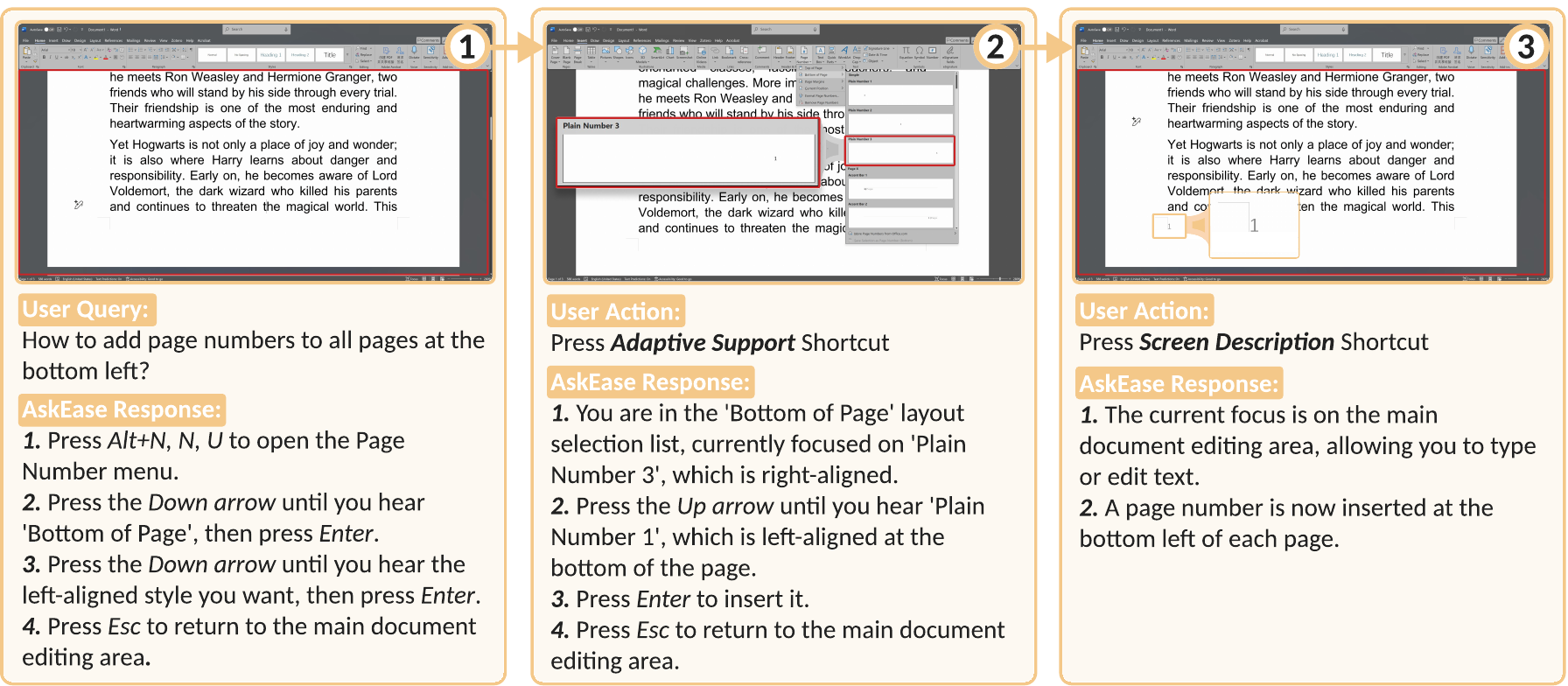}
    \caption{Multi-turn assistance in Word: (1) initial step-by-step guidance for inserting left-aligned page numbers, (2) adaptive guidance to \textbf{address an accessibility issue} where the left-aligned page number style is incorrectly displayed as ``Plain number 1'', and (3) confirmation of correct page number placement.
}
    \Description{Three-panel sequence: Panel 1 shows Word document with user query "How to add page numbers to all pages at the bottom left?" and AskEase response providing four-step guidance: press Alt+N, N, U to open Page Number menu, press Down arrow until hearing "Bottom of Page" then press Enter, press Down arrow until hearing left-aligned style then press Enter, and press Esc to return to main document editing area. Panel 2 displays document with "Bottom of Page" layout selection list visible, user action "Press Adaptive Support Shortcut", and AskEase response explaining current focus is on "Bottom of Page" layout selection list with "Plain Number 3" right-aligned, instructing to press Up arrow until hearing "Plain Number 1" which is left-aligned at bottom of page, then press Enter to insert and Esc to return to document. Panel 3 shows a document with the user action "Press Screen Description Shortcut" and AskEase response confirming that the current focus is on the main document editing area, and the page number is now inserted at the bottom left of each page.}
    \label{fig:word}
\end{figure*}
Nancy is a blind student who is writing a paper in Word and wants to format it according to the guidelines. 
However, she is unfamiliar with document formatting in Word. She invokes \emph{Contextual Q\&A} and asks, ``How to add page numbers to all pages at the bottom left?'' (\autoref{fig:word} (1)). The assistant provides clear, step-by-step instructions. Following the guidance, Nancy navigates Word's ribbon interface using keyboard shortcuts and locates the page number options.
When she tries to select the desired page number format, she hears only ``page number 1,'' ``page number 2,'' ``page number 3,'' with no indication of a ``left-aligned style.'' To resolve this, she invokes \emph{Adaptive Support}. The assistant analyzes the current screen state and recent SR traces, generating refined instructions that clarify which option corresponds to the left-aligned style (\autoref{fig:word} (2)). Nancy follows the updated steps and successfully selects the left-aligned page number.
To verify that the page numbers have been correctly inserted in the footer, Nancy invokes \emph{Screen Description}. The assistant confirms that page numbers appear at the bottom left of all pages (\autoref{fig:word} (3)).
With the assistant, Nancy completes tasks efficiently, understands visual elements through verbal descriptions, and interacts independently with software features.

\subsection{Robustness and Computational Cost}
\label{sec:robustness-evaluation}
We evaluate \systemname{} with the state-of-the-art OpenAI \texttt{GPT-5}~\cite{openai_gpt5}, selected for its reasoning and planning abilities that support complex task handling. 
\nan{We also experimented with models such as GPT-4o and Gemma 3~\cite{team2025gemma}, but found that GPT-5 achieved higher accuracy in on-screen content recognition and showed better adherence to knowledge context. Therefore, we selected GPT-5 for our robustness testing and user study. Importantly, our approach is model-agnostic: as MLLMs continue to advance, \systemname{} can be adapted to other local models or more advanced models.}

To this end, we sampled 45 tasks from the Windows Agent Arena dataset~\cite{bonatti2024windows}, a benchmark covering six computer use scenarios: web browsing, office productivity, Windows system operations, coding, media and video, and general Windows utilities. 
These tasks are multi-step; a \texttt{GPT-4V-based} UI automation agent achieves only a 19.5\% success rate on this dataset~\cite{bonatti2024windows}.
The dataset spans 12 applications (e.g., Clock, VS Code, and Edge). 
We targeted up to four tasks per application, but scarcity in some cases yielded 45 tasks in total \fix{(see supplementary materials Table A)}.

For each task, we first ensured that NVDA focus was placed on the application's main interface (or on the desktop if the task required launching the application). The \emph{Contextual Q\&A} shortcut was then invoked to open \systemname{}.  
The task description was entered verbatim, and the task was attempted by following the generated step-by-step guidance using only SR interaction. 
A task was marked successful if it was fully completed. 
When a step was infeasible, \emph{Adaptive Support} was invoked (up to three times per task) to request adaptive instructions and examine the model's capacity to adapt its guidance based on the given context.
\nan{Similar to prior work~\cite{ehtesham2025topsen}, two authors with beginner-level experience using NVDA conducted the evaluation with the display turned off, relying solely on audio feedback and keyboard interactions to perform the evaluation tasks.}

Of the 45 tasks, 18 were completed after a single query (40.0\%). 
With up to three \emph{Adaptive Support} rounds, an additional 25 tasks were completed, with an average of 1.80 rounds ($\sigma=0.73$), yielding an overall success rate of 96.6\%. 
The remaining two failures included tasks such as ``Open Paint and draw a red circle,'' where the guidance required mouse dragging, which was not achievable with keyboard-only interaction.
The average latency per \emph{Contextual Q\&A} invocation was 10.06 s ($\sigma=1.36$), with an average API cost of \$0.0050 USD ($\sigma=0.0017$). The average input and output token lengths were 2344 ($\sigma=901$) and 204 ($\sigma=54$), respectively. 
For \emph{Adaptive Support}, the average latency was 8.18 s ($\sigma=1.14$) with an average API cost of \$0.0046 USD ($\sigma=0.00057$). The average input and output token lengths were 1848 ($\sigma=77$) and 229 ($\sigma=47$).


\section{User Study}
\label{sec-evaluation}

We conducted a within-subject study with 12 SR users to evaluate the effectiveness of \systemname{} compared to their usual tools (e.g., search engines or AI assistants). Specifically, we aim to answer the following research questions:

\begin{enumerate}[label=\textbf{RQ\arabic*.}]
    \item How does \systemname{} affect SR users' task completion performance?
    \item How do SR users leverage \systemname{} to overcome challenges in computer use?
    \item How does \systemname{} influence SR users’ perceived workload during tasks?
    \item How do SR users evaluate the overall usability and helpfulness of \systemname{} in supporting computer interactions?
\end{enumerate}




\subsection{Participants}
We recruited 12 participants (P1-P12; 6 female, 6 male), all of whom were SR users with blindness, with ages ranging from 21 to 44 ($M = 32$, $\sigma = 6.24$).
Recruitment was conducted through professional networks and community outreach, \nan{targeting diversity in occupation, gender, and age to better reflect the range of SR users. To better observe how users seek assistance with an unfamiliar complex interface, we intentionally recruited participants who had prior exposure to the software but were unfamiliar with it.} 
All participants had prior experience using \nan{LLM tools} (e.g., ChatGPT, Gemini): two were \nan{highly experienced, able to use LLM tools for complex tasks and skilled in advanced prompting and tool features}; nine \nan{had moderate experience, mainly engaging in ordinary chat-based interactions} for tasks such as describing images, acquiring general knowledge, and writing; and one had \nan{low experience, with little hands-on practice}.
Participants provided informed consent and received \$30 USD (or local equivalent) for their participation. 
Participant demographics are summarized in ~\autoref{tab:participants}.


\begin{table*}[htbp]
\centering
\caption{Demographics of participants: \nan{LLM Tool} Experience refers to self-reported experience with \nan{LLM tools}, and Software Usage refers to self-reported experience with the software used in the study; both are categorized as high, medium, or low.}
\Description{A six-column table listing demographic information for twelve participants in the user study. Columns include participant ID, Occupation, Gender, Age, LLM Tool experience, and Software Usage. Occupations include roles such as entrepreneur, developer, massage therapist, and writer.}
\begin{tabular}{c c c c c c}
\toprule
ID & Occupation & Gender & Age & \nan{LLM Tool} Experience & Software Usage \\
\midrule
1  & Entrepreneur        & M & 33 & High   & Medium \\
2  & Developer           & M & 25 & High   & Low    \\
3  & Content Creator     & M & 44 & Medium & Low    \\
4  & Massage Therapist   & F & 26 & Medium & Low    \\
5  & SR Customer Support & F & 38 & Medium & Low    \\
6  & Massage Therapist   & F & 28 & Medium & Low    \\
7  & Music Performer     & M & 32 & Low    & Low    \\
8  & Piano Tuner         & M & 27 & Medium & Medium \\
9  & Game Planner        & M & 34 & Medium & Low    \\
10 & Student             & F & 21 & Medium & Low    \\
11 & Massage Therapist   & F & 33 & Medium & Low    \\
12 & Writer              & F & 38 & Medium & Low    \\
\bottomrule
\end{tabular}
\label{tab:participants}
\end{table*}

\subsection{Choice of Applications and Tasks}
\label{sec:user-study-app-task}
We selected Microsoft Word and Microsoft Excel for our study, as these productivity tools are commonly used by SR users but remain challenging to navigate and operate efficiently~\cite{perera2025sky}. 
For each application, we designed two task sets (A and B), each containing two tasks. To minimize learning effects, the two sets were constructed to avoid highly similar operations. Tasks covered typical operations, such as document formatting and numerical calculations, but incorporated specific features or options that were likely unfamiliar to participants. 
This design reflects real-world computer use, where users must continually explore and experiment with new features to expand their skills and accomplish a broader range of tasks.
Selecting unfamiliar tasks also allowed us to observe the challenges participants encountered when learning to use new features and how they sought help in the process.
Task designs were adapted from Windows Agent Arena~\cite{bonatti2024windows} and adjusted to ensure comparable difficulty across tasks. 
We controlled for task difficulty by ensuring that each task contained a similar number of steps, thereby enabling fair within-subject comparisons.



\subsection{Procedure}
Each study session was conducted remotely and lasted approximately 80 minutes. Prior to the experiment, participants installed \systemname{} and provided basic demographic and background information.
During the study, each participant was randomly assigned to one application (Word or Excel) to control for variance in experience due to prior familiarity. Within the chosen application, they completed two task sets under different conditions---using \systemname{} or their usual tools (e.g., search engines or AI assistants) as a baseline.
To counterbalance potential order effects, we employed a 2 × 2 Latin Square randomized crossover design within each application, balancing the order of conditions and task sets. 
Before using \systemname{}, participants completed a brief tutorial introducing its main features.
During task completion, participants were encouraged to think aloud, verbalizing their thoughts, strategies, and challenges. Each task set had a 20-minute time limit, which participants were not informed of in advance to avoid creating performance pressure. When the time was reached, participants were gently notified and could choose whether to continue or stop.

After completing each task set, participants completed the NASA-TLX questionnaire~\cite{Sandra1988NASA-TLX} to assess perceived cognitive workload. For tasks completed with \systemname{}, participants also responded to a 7-point Likert-scale survey (see~\autoref{fig:user_satisfaction}). Finally, semi-structured interviews were conducted to gather qualitative feedback on participants' experiences and suggestions for improvement.
The study protocol was approved by the Institutional Review Board (IRB).

\subsection{Data Analysis}
With participants' consent, we recorded screen and audio, logged interactions with \systemname{}, and collected questionnaire responses.
For task completion, we counted only the tasks completed correctly within 20 minutes to ensure a fair comparison.
Each measure (task completion and NASA-TLX) was analyzed as a continuous outcome using linear mixed-effects models~\cite{lindstrom1988newton} with a participant-level random intercept.
Fixed effects included task condition (using \systemname{} or baseline tools), task (software × task group), and task order. Model comparison followed standard practices using the Bayesian Information Criterion (BIC)~\cite{neath2012bayesian}, with the Akaike Information Criterion (AIC)~\cite{sakamoto1986akaike} and likelihood ratio tests (LRT)~\cite{anisimova2006approximate} for nested models, and indicated that differences in scores were primarily explained by task condition. All subsequent analyses were conducted using the BIC-selected model.
We analyzed the interaction logs to calculate the frequency of features used. Recorded audio was transcribed and examined to address the research questions.

\section{Study Results}
\label{sec-result}

In the within-subject user study, \systemname{} significantly increased task success while reducing perceived workload. Below, we present the results corresponding to the four research questions~\nan{and investigate unideal cases}.


\subsection{Task Completion (RQ1)}



\textbf{\systemname{} facilitated more successful completion of unfamiliar computer tasks.} 
When using \systemname{}, participants completed on average 1.5 of 2 tasks (all completed at least one), significantly more than the 0.5 tasks they completed with baseline tools ($\beta = 1.00 \pm 0.26$, $p < 0.001$; \autoref{fig:nasa_tlx}).
\nan{Six participants completed both tasks within the 20-minute task period. In contrast, only two participants completed both tasks when using the baseline tools, and three chose to stop early due to increasing pressure (See \autoref{tab:participants task performance}).}

\begin{table*}[htbp]
    \centering
    \caption{\fix{Comparison of task completion time and number of tasks completed across participants using \systemname{} versus baseline tools. * indicates that the participant abandoned the task before the allotted 20 minutes.}}
    \Description{The table compares task performance between the AskEase and Baseline Tools conditions for 12 participants. Columns include participant ID, completion time, and the number of tasks completed under each condition. It shows that users with AskEase generally completed tasks faster, accomplished more tasks, and did not abandon any tasks.}
    \begin{tabular}{ c c c c c }
    \toprule
    \multirow{2}{*}{\textbf{ID}} & \multicolumn{2}{c}{\textbf{\systemname{}}} & \multicolumn{2}{c}{\textbf{Baseline Tools}} \\
    \cmidrule(lr){2-3} \cmidrule(lr){4-5}
     & \textbf{Completion Time} & \textbf{\#Tasks Completed} & \textbf{Completion Time} & \textbf{\#Tasks Completed} \\
    \midrule
    1 & 14:12 & 2 & \timeout{>20:00} & \fail{0} \\
    2 & \timeout{>20:00} & 1 & 07:55 & 2 \\
    3 & \timeout{>20:00} & 1 & \timeout{>20:00} & \fail{0}\\
    4 & \timeout{>20:00} & 1 & \gaveup{2:13*} & \fail{0}\\
    5 & 19:31 & 2 & \timeout{>20:00} & 1 \\
    6 & 16:07 & 2 & \timeout{>20:00} & \fail{0}\\
    7 & 17:39 & 2 & 18:02 & 2 \\
    8 & \timeout{>20:00} & 1 & \timeout{>20:00} & \fail{0}\\
    9 & 19:54 & 2 & \timeout{>20:00} & \fail{0}\\
    10 & \timeout{>20:00} & 1 & \gaveup{16:30*} & \fail{0}\\
    11 & \timeout{>20:00} & 1 & \gaveup{16:23*} & 1 \\
    12 & 19:26 & 2 & \timeout{>20:00} & \fail{0}\\
    \bottomrule
    \end{tabular}
    \label{tab:participants task performance}
\end{table*}

\nan{Participants used different baseline tools to complete their tasks.} Three participants sought online tutorials on search engines, while eight attempted to use chat-based AI assistants (e.g., Gemini) for guidance. One participant with programming experience (P2) employed a coding agent (Qwen Code~\cite{qwen_code}) to automate Excel tasks via Python scripting.
Among those who used on search engines or AI tools,
two preferred interfaces designed for SR users, featuring minimalistic layouts and straightforward navigation. P3 noted that these tools were convenient because a single shortcut provided immediate access to the search box, whereas conventional browsers required more complex interactions.

\subsection{Feature Usage (RQ2)}
\nan{\systemname{} was essential for task completion. All participants worked closely with the system, generally following its instructions while occasionally applying their own strategies (e.g., P7 used arrow-key navigation instead of the suggested shortcut).}
%
Participants engaged most frequently with \textit{Contextual Q\&A}, using it an average of 5.25 times ($\sigma = 2.55$). P2, who used this feature 11 times---the highest among participants---described it as particularly convenient, noting that it ``\emph{allows me to ask questions effortlessly and get directly to the solution.}''
\nan{Based on our observations, participants' questions fell into four types (\autoref{tab:contextQA-categories}). \textit{Task-level questions} focus on the overall goal and are often expressed by copying or rephrasing the task instructions. \textit{Step-level questions} focus on a single step within a task and often reflect participants' mental models for decomposing tasks.
When encountering uncertainty or obstacles, participants occasionally turned to \textit{status confirmation questions} to check the current interaction state, or \textit{troubleshooting questions} to figure out why a step could not be completed. These latter two types resembled the \textit{Screen Description} and \textit{Adaptive Support} features, but their own questions were more flexible and targeted, reflecting their individual strategies, interaction habits, and needs.}

\textit{Adaptive Support} was the next most frequently used feature (M = 2.17, $\sigma = 1.86$) and provided targeted assistance when participants encountered difficulties. P5, for example, invoked \textit{Adaptive Support} six times during a single task. When she struggled to format text in uppercase, the feature suggested an alternative approach, enabling her to continue. She appreciated, ``\emph{I don't have to switch between windows and can stay focused while exploring new functions.}''

\nan{After sending a request, participants waited and followed the system’s audio cues to track progress.} To review responses, participants frequently used \textit{Previous Step} (M = 19.83, $\sigma = 17.97$) and \textit{Next Step} (M = 29.08, $\sigma = 23.57$).
P3 used \textit{Previous Step} and \textit{Next Step} heavily (15 and 25 times, respectively). He explained, ``\emph{I really love this feature---it makes it so easy to review the AI's responses.}''
Overall, these findings illustrate how participants leveraged different features of \systemname{} for immediate problem-solving, adaptive guidance, and structured navigation, thereby facilitating more effective interaction with unfamiliar software.

\begin{table*}[htbp]
\centering
\caption{\fix{Participants’ question types and representative examples.}}
\Description{The table summarizes four categories of user questions: task-level query (24 occurrences, e.g., “How can I convert CSV content in a document into a table?” asked by P9), step-level query (27 occurrences, e.g., “How to insert a column before this one?” asked by P3), status-confirmation (7 occurrences, e.g., “Is my current operation correct?” asked by P10), and troubleshooting (5 occurrences, e.g., “Why can’t E2 calculate the age correctly?” asked by P3). For each category, the table provides representative examples with participant IDs and total frequency.}
\label{tab:contextQA-categories}
\begin{tabular}{l p{11cm} c }
\toprule
\textbf{Question Type} & \textbf{Examples} & \textbf{Count} \\
\midrule
\multirow{4}{*}{task-level}& 
$\cdot$ I am a beginner in Excel, so I need your help. I want to count the number of people for each gender according to the ``Gender'' column in the table, and record the results in a new table with the headers ``Gender'' and ``Count''. (P2) \newline
$\cdot$ How can I convert CSV content in a document into a table? (P9)
& 24 \\
\midrule
\multirow{2}{*}{step-level}& 
$\cdot$ How to insert a column before this one? (P3)
\newline
$\cdot$ How do I select text using NVDA? (P5) 
& 27 \\
\midrule
\multirow{2}{*}{\makecell[l]{status \\ confirmation}}& 
$\cdot$ Am I currently in the Pivot Table editing interface? (P2) \newline
$\cdot$ Is my current operation correct? (P10) 
& 7 \\
\midrule
\multirow{2}{*}{troubleshooting}& 
$\cdot$ Why can't E2 calculate the age correctly? (P3) \newline
$\cdot$ Ctrl+1 seems to be used by other software. How can I open the settings dialog? (P2)
& 5 \\
\bottomrule
\end{tabular}
\end{table*}

\subsection{Perceived Workload (RQ3)}

\begin{figure*}[htbp]
    \centering
    \includegraphics[width=\linewidth]{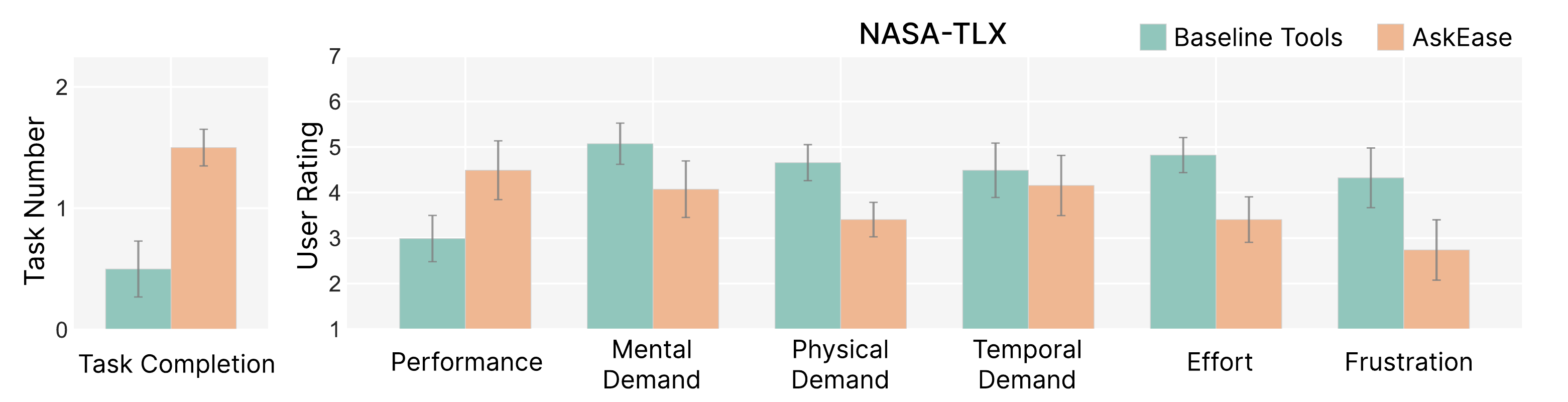}
    \caption{Comparison of task completion and NASA-TLX ratings between baseline tools and \systemname{}. NASA-TLX scores range from 1 (strong disagreement) to 7 (strong agreement). Higher scores indicate higher perceived workload, except for performance, where higher values reflect better perceived task performance.}
    \Description{Two sets of bar charts compare task completion and user ratings under two conditions: "Baseline Tools" (green bars), representing commonly used screen reader-compatible tools, and "AskEase" (orange bars), representing our proposed system. The left chart shows task completion counts, with higher completion observed using AskEase. The right chart displays NASA-TLX ratings across six dimensions: Performance, Mental Demand, Physical Demand, Temporal Demand, Effort, and Frustration. Performance is rated higher with AskEase, while workload-related dimensions show lower ratings, indicating reduced perceived workload. Error bars indicate data variability.}
    \label{fig:nasa_tlx}
\end{figure*}

\textbf{Participants reported that \systemname{} significantly lowered their workload.}
As shown in~\autoref{fig:nasa_tlx}, participants perceived their workload to be lower when using \systemname{} compared to baseline tools.
Quantitatively, \systemname{} significantly reduced \textit{Physical Demand} ($\beta = -1.25 \pm 0.52$, $p < 0.05$), \textit{Effort} ($\beta = -1.42 \pm 0.60$, $p < 0.05$), and \textit{Frustration} ($\beta = -1.58 \pm 0.63$, $p < 0.05$). At the same time, perceived \textit{Performance} was significantly improved ($\beta = +1.50 \pm 0.55$, $p < 0.01$).  
Together, these results indicate that \systemname{} provided participants with a more favorable experience.


\textbf{Participants faced challenges when seeking help with baseline tools.} 
They described the tutorials or AI-generated responses as ``\emph{overly verbose}'' (e.g., P5, P6, P8), ``\emph{hard to navigate}'' (e.g., P3, P7, P12), and ``\emph{not friendly to screen reader use}'' (e.g., P6, P7, P12).
For example, P3 found tutorials via keyword search, but they contained many screenshots and mouse-based instructions. He reported, ``\emph{I have to translate these instructions into my own experience.}'' 
P5 asked an LLM-based assistant for guidance, but since her prompt did not specify the application, the assistant provided instructions for three or four different text editing tools.
P10 added, ``\emph{Finding a suitable tutorial is not easy.}''  
These observations are consistent with prior work~\cite{perera2025sky,chen2025screenreaderusersvibe} and highlight the need for an on-demand, context-aware assistant.




\subsection{User Perceptions (RQ4)}
\begin{figure*}[ht]
    \centering
    \includegraphics[width=0.9\linewidth]{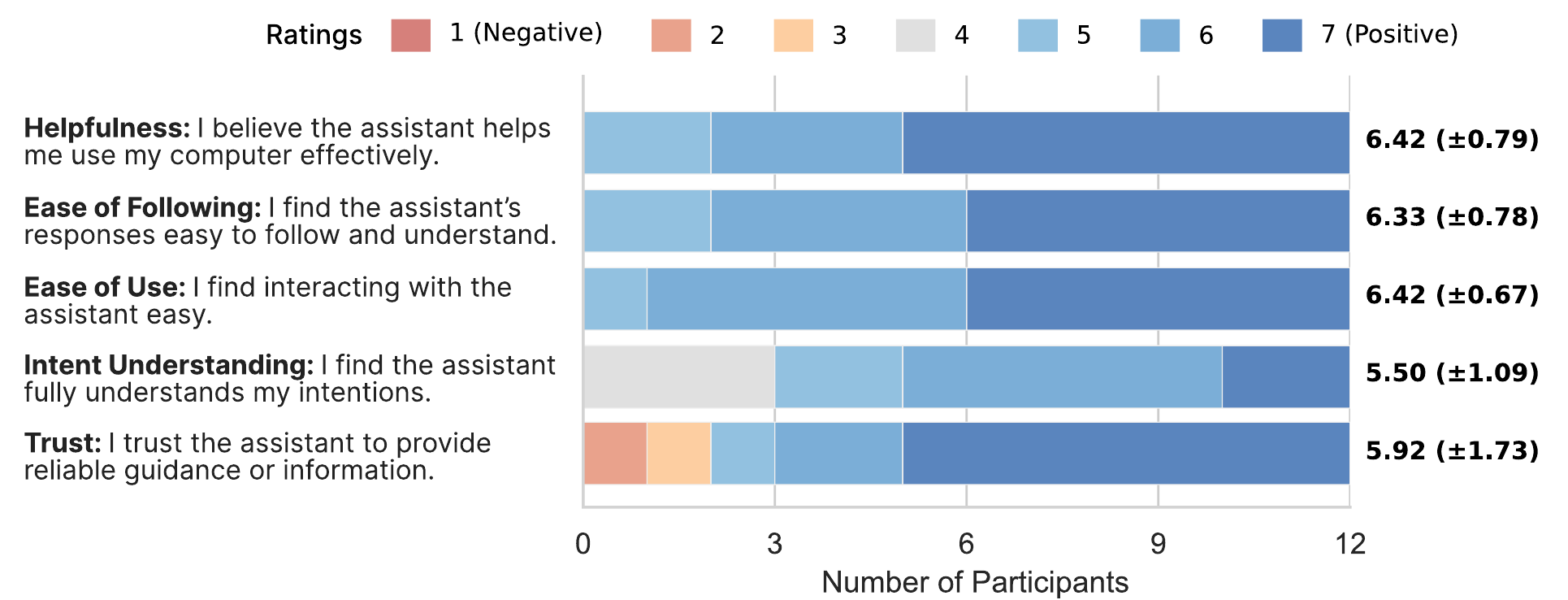}
    \caption{Participants' responses to the post-task 7-point Likert scale survey. Higher scores indicate stronger agreement.}
    \Description{Stacked horizontal bar chart illustrating user satisfaction with AskEase across five dimensions, rated on a 7-point Likert scale. The dimensions include: Helpfulness: I believe the assistant helps me use my computer effectively. Ease of Following: I find the assistant’s responses easy to follow and understand. Ease of Use: I find interacting with the assistant easy. Intent Understanding: I find the assistant fully understands my intentions. Trust: I trust the assistant to provide reliable guidance or information. Most ratings cluster toward the positive end (scores 5–7). Helpfulness and Ease of Use received the highest average scores (6.42), followed by Ease of Following (6.33). Intent Understanding averaged 5.50. Trust had a relatively high average (5.92) but showed greater variability, with some ratings as low as 2 or 3. Standard deviations ranged from 0.67 to 1.73, with Trust showing the highest variability. Overall, the chart indicates high satisfaction, with trust and intent understanding being more mixed.}
    \label{fig:user_satisfaction}
\end{figure*}

Overall, participants found that the system provided helpful guidance that could support them in future computer use ($M = 6.42$, $\sigma = 0.79$), and appreciated its user-friendly interaction and easy-to-follow responses (see \autoref{fig:user_satisfaction}).

\subsubsection{Screen Reader User-Friendly Responses.}
Participants rated \systemname{} highly for \textit{Ease of Following} ($M = 6.33$, $\sigma = 0.78$), finding its responses more accessible than those of general AI assistants in terms of wording and structure.

\textbf{Participants highlighted that \systemname{} consistently recommended \emph{non-visual description and keyboard operations}}. 
Baseline AI assistants or search engines often returned visual instructions (e.g., ``click the button'') and rarely offered keyboard shortcuts unless explicitly prompted. 
During the experiment, only two participants noted that they were using a SR or requested non-visual alternatives, which often resulted in unsuitable responses. 

\textbf{Participants appreciated \systemname{}'s \emph{step-by-step guidance}, which helped them follow instructions in a straightforward way.}
P6 explained, ``\emph{General AI assistants often delivered long, unstructured replies that were difficult to follow. \systemname{} offered concise, actionable instructions---one step, one operation.}''
Similarly, P3 noted, ``\emph{For tasks broken down step by step, I can proceed gradually, which helps me maintain a clear understanding of what to do next.}''

\subsubsection{Seamless Interaction with \systemname{}.}

Participants rated \systemname{} highly for \textit{Ease of Use} ($M = 6.42$, $\sigma = 0.67$), noting that it provided assistance seamlessly and supported a smooth, efficient interaction.

\textbf{Situation Awareness enabled seamless question asking, eliminating the need to copy and paste.}
P1 described, ``\emph{In the past, when I encountered difficulties, I would take a screenshot and then send it to the AI assistant. This process was cumbersome. It can see my screen, which is great.}''
As P5 explained, ``\emph{When using other AI assistants, I often had to switch back and forth, copying task content and describing what I needed to do. I would forget what I was supposed to do when switching. This system allows me to complete tasks smoothly and continuously.}''

\textbf{Reduced window switching and focus shifts made the interactions more accessible for SR users.}
Participants frequently used the \textit{Previous Step} and \textit{Next Step} features, as well as \emph{Adaptive Support}, without opening new windows or losing focus. 
P10 said, ``\emph{I often need to go back and forth to check previous steps. The shortcuts are very convenient; I don't have to switch windows, which saves time and effort, and I don't have to worry about losing my focus.}''

\subsubsection{Intent Understanding.}
\label{sec:intent}
Participants generally agreed that\linebreak\systemname{} could accurately capture their intentions ($M = 5.50$, $\sigma = 1.09$). 
Analysis of the interaction logs further revealed that, in some cases, participants---such as P11---did not provide fully clear task descriptions; nonetheless, \systemname{} was able to generate appropriate responses by inferring from the available context. However, some participants also noted instances where their intentions and the assistant's interpretations did not fully align. We further analyzed these cases to identify the underlying causes.

\textbf{Unclear prompts as a barrier.}
When users were unfamiliar with a task or software, they often struggled to clearly articulate their goals, which in turn challenged the assistant's intent recognition. 
For instance, P10 attempted to format an \texttt{ID} column in Excel into 7-digit numbers with leading zeros. 
She asked, ``\emph{I have rows of numbers. How can I make all of them start with 0?}'' 
Because she mistakenly described a column as ``\emph{rows}'' and used the ambiguous phrasing ``\emph{start with 0,}'' the assistant interpreted her request as simply prepending a single zero to each number. 
It shows how ambiguity in user prompts---often rooted in unfamiliarity with domain-specific terminology---can propagate into misinterpretations. 
%

\textbf{Misalignment with individual preferences.}
Some users felt their intent was not fully met due to mismatches in preferred response styles. 
While many valued concise answers that directly addressed their questions, others desired broader solution spaces. 
As P8 explained, ``\emph{The assistant usually gives me one solution, but I want to see all possible solutions. Having a more comprehensive view helps me learn better.}''
Similarly, P4 noted that some responses were perceived as indirect, which made it challenging to execute shortcut instructions like ``\texttt{Alt+H, N}'' without guidance that the key presses should be performed sequentially. 
While this phrasing aligns with accessibility conventions, it highlights a mismatch between standard instructions and the user's expectations or familiarity.

\subsubsection{Trust as a Gradual Process}
\label{sec:trust}

Trust in \systemname{} varied across participants (\autoref{fig:user_satisfaction}). 
A closer analysis revealed that these differences were influenced by prior experiences with AI tools, familiarity with the system, and individual strategies for interacting with AI. 
These findings suggest that trust develops gradually through repeated positive interactions, with consistent and reliable outputs helping users make informed decisions.

\textbf{Initial skepticism and trust accumulation.}  
Trust appeared to develop incrementally over time, shaped by past experiences and familiarity with AI systems. 
Some participants reported initial skepticism due to previous encounters with hallucinations or inconsistent AI responses, often requiring multiple follow-up questions to verify reliability. 
P8 explained in detail, ``\emph{In my previous experiences with AI assistants, I was sometimes misled by incorrect responses, which made me cautious about relying on them. When I first used \systemname{}, I did not immediately trust the suggestions. Over time, as I observed consistent and helpful responses, I began to feel more confident and gradually developed trust in the system.}''

\textbf{Maintaining control while using AI.}  
Other participants emphasized the importance of exercising caution. 
P3 said, ``\emph{I typically combine AI suggestions with my own judgment rather than following them blindly. I evaluate whether the generated steps align with my understanding and sometimes adjust the sequence based on my knowledge. This approach allows me to benefit from the AI's guidance while maintaining control over the outcome.}''

\subsubsection{Value for Future Use}

Participants highlighted that \systemname{} has the potential to lower barriers to learning, address accessibility challenges, and its potential for career development and broader social empowerment. 
As P12 described, using the tool was ``\emph{like discovering a new world}'', illustrating \systemname{}'s value for future use.

\textbf{Supporting learning and digital skill development.}
Participants described \systemname{} not only as an assistant for immediate tasks but also as a companion for acquiring new competencies.
P10 reported,
``\emph{With guidance through a single run of the operation, I can learn the essential steps and logic. If I encounter a similar problem later, I will be able to solve it myself.}'' 
Similarly, P3 recounted his repeated attempts to learn VS Code, which he ultimately abandoned because it was too difficult to navigate: ``\emph{I had tried installing VS Code many times but gave up each time because it was too hard to get started. After experiencing this assistant, I feel I can ask it whenever I am unsure. It provides feasible guidance and makes learning a new application feel less daunting.}''

\textbf{Overcoming critical barriers in everyday use.}
Participants reflected on the accessibility challenges they frequently encounter and discussed how \systemname{} could help address them. 
P5, who previously provided support for SR users, noted that many questions about SR use could be answered more quickly with such an assistant.
Several participants (P2, P6, P9, and P12) highlighted handling verification codes as a particularly persistent challenge. 
These difficulties illustrate opportunities for \systemname{} to provide on-demand guidance, enabling users to navigate tasks that are inaccessible.
P2 elaborated on the system's potential, noting, ``\emph{It could not only recognize text, but also guide me in situations that are otherwise difficult to navigate non-visually, making tasks that are currently inaccessible much more approachable.}''


\textbf{Autonomy and Professional Growth.}
Participants emphasized \systemname{}'s potential to promote personal independence, support career development, and foster social empowerment.  
Participants (e.g., P9, P12) reported that they had often relied on others in the past but hoped to become more independent. They perceived that \systemname{}'s on-demand assistance could help them achieve this autonomy.
In addition, P5 noted, ``\emph{This tool can help me adapt to and learn new software and technologies, which is very important for career development and staying competitive.}'' 
P8 further reflected on how digital accessibility expands professional opportunities, ``\emph{With accessible digital tools, people with visual impairments can engage in a wider range of careers. In the past, we often worked in traditional roles like massage, but now more individuals are beginning to explore programming and other digital work. During the study, I thought that if I could better understand Word formatting, it would reduce gaps with other colleagues and allow for fairer access to opportunities.}''
\subsection{\nan{Unideal Cases and Coping Strategies}}
\label{sec:unideal}
\nan{In addition to user-related factors, such as ambiguous questions or mismatched personal preferences (\autoref{sec:intent}), we observed that several system-related limitations. In this section, we analyze unideal cases and discuss how they were addressed through user experience and adaptive support.}

\nan{First, though we collected three types of \emph{environment context}, it was sometimes insufficient.
For instance, P2 was suggested by the assistant to use the shortcut ``Ctrl+1'' to open formatting settings. However, this shortcut conflicted with shortcuts in other software, preventing further progress. In another case, P3 attempted to calculate ages in Excel. However, \systemname{} suggested a formula referencing the wrong cell, likely due to the small cell size and low resolution. Such incomplete or coarse-grained environment information can prevent \systemname{} from providing accurate guidance.}

\nan{Second, while we provided the assistant with \emph{knowledge context} related to the user's question through RAG, the documentation was not always comprehensive. Consequently, the model sometimes hallucinated, generating shortcuts or instructing users to interact with elements that did not exist. For example, during robustness evaluation, when attempting to enable the ``Do Not Track'' feature in Chrome, \systemname{} provided navigation instructions that included non-existent menu paths and incorrect keyboard shortcuts.}

\nan{
Some of these unideal cases could be addressed by users' prior experience. For example, P2 recognized the potential shortcut conflict and included this information in a follow-up prompt, allowing the assistant to provide a more appropriate suggestion. Others could be addressed via \textit{Adaptive Support}. 
When a user cannot proceed, \systemname{} analyzes the ambiguous or incorrect step and provides targeted guidance or alternative solutions, helping the user continue successfully.
Nevertheless, since certain information may not appear in training data or knowledge context, LLM-based assistants cannot guarantee correct guidance in all situations. We therefore advocate that developers adhere to accessibility guidelines and provide accessible documentation.}

\section{Discussion}
\label{sec-discussion}


\subsection{The Need for Personalization}

Our study indicates that personalization is crucial for effectively supporting diverse SR users, consistent with prior research~\cite{danmuA11y2025,perera2025sky,Customization2024}.
Participants exhibited varying preferences in guidance style, content detail, interaction habits, and task priorities, as follows:

\textbf{Varied guidance depth based on user proficiency.} Although SR users generally preferred concise wording~\cite{venkatraman2024you}, guidance still needs to be adapted to their experience level, providing an appropriate degree of detail.
For example, P4 was unsure whether the keyboard shortcut ``\texttt{Alt+T, N}'' should be pressed simultaneously or in sequence, showing that even simple instructions can be confusing without clarification. 
Tailored modes---such as ``beginner'' and ``advanced''---could allow users to select the appropriate level of explanation for their expertise.

\textbf{Diverse expectations for content breadth.} While many users prefer direct guidance, some favored instructions that present multiple alternative solutions, allowing them to select the approach best suited to their needs while also promoting cognitive engagement and facilitating the acquisition of new skills (e.g., P1, P8). Supporting flexible content breadth can accommodate both users seeking efficiency and explorers aiming for a broader understanding.

\textbf{Diverse interaction preferences.} Users differ in how they review system responses. Although the sentence-level review shortcut provided by \systemname{} was widely adopted, some participants (e.g., P9) still preferred to open the dialogue and navigate word by word. Designing shortcuts with these differences in mind, and offering more flexible navigation options, can better accommodate diverse interaction preferences.

\textbf{Efficiency vs. Accuracy.} Users' focus on efficiency versus accuracy shapes their expectations. Some users (e.g., P1) place a higher priority on response speed, whereas others (e.g., P12) value reliability, using the system's slower but higher-quality RAG method. 
\nan{This reflects a trade-off between response time and accuracy, as faster outputs may reduce correctness.}
AI systems that recognize and adapt to these priorities can better support diverse workflows.

These diverse preferences highlight the importance of accounting for variability in users' experience levels, individual characteristics, and interaction styles when designing AI assistance. A promising direction is \textbf{adaptive personalization}, in which assistants leverage contexts---such as users' prior computer operations and previous interactions with the assistant---to dynamically adjust the level and style of guidance. Such context-aware adaptation can better accommodate heterogeneous needs and preferences.

\subsection{Rethinking Step-by-Step Guidance in the Era of Increasing Automation}

Our study revealed an important observation that calls for reflection on the role of step-by-step guidance in an era of increasingly automated tools. 
In one case, P2 completed an Excel task entirely through a coding agent without even opening the Excel application. 
This raises a critical question: \emph{Do users still need step-by-step guidance when agent automation can accomplish tasks directly?}
We discussed its potential directions:

\textbf{Even automated tools require learning.}
Given the above example, P2 was familiar with the coding agent because he is a programmer with frequent exposure to cutting-edge technologies. In contrast, other participants reported that engaging with novel tools often posed significant challenges~\cite{chen2025screenreaderusersvibe,adnin2024look,flores2025impact}. Step-by-step guidance can lead the learning process for these users, helping them gain familiarity with emerging automated tools.

\textbf{Automated tools can be risky and require users' oversight.}
During P2's task, the agent unintentionally converted the Excel file into an empty file, causing an error. 
He was only able to continue after manually duplicating the file. This example demonstrates that automated tools do not always complete tasks perfectly and can introduce errors or additional risks~\cite{ayyamperumal2024current,wang2025survey}, and that users need to develop the ability to intervene and recover from such outcomes~\cite{mcbride2014understanding,shneiderman2020human}.
It highlights the value of guided step-by-step interaction as a safeguard~\cite{tang2024prioritizing}.

\textbf{Step-by-step guidance supports deeper understanding and exploration.}
While automation may improve efficiency for completing tasks, participants expressed a desire to understand underlying processes and explore software functionality~\cite{carroll1987paradox,kiani2019beyond}. 
Structured guidance not only aids task completion but also satisfies users' needs for explainability, fostering deeper engagement and skill development~\cite{khurana2025me,khurana2024and}.

Overall, these observations suggest that step-by-step guidance remains important even in highly automated environments. It can support users across different levels of expertise, mitigate risks associated with automation, and fulfill explainability and exploratory needs that purely automated solutions may overlook.

\subsection{\nan{AI Assistance Supports, but Accessibility by Design Prevails}}

\nan{Our study shows that \systemname{} provides SR users with context-aware, on-demand support, helping them explore new tools and navigate complex applications. Participants appreciated having an assistant that can guide task execution and confirm on-screen information, facilitating learning and interaction with software.}

\nan{At the same time, we learn the lesson that AI assistants cannot replace good accessibility design. \systemname{} tends to be more effective on tools that follow SR conventions and provide accessibility support, but its performance can be limited on less accessible applications.
For instance, unlabeled interface elements can reduce response reliability, as the model may hallucinate when inferring their purpose~\cite{adnin2024look,alharbi2024misfitting}. Moreover, some issues are inherently out of scope, such as elements that are not operable via keyboard navigation.
Previous studies on AI-powered ``overlays'' highlight the risks of surface-level fixes, which can obscure underlying accessibility problems and reduce developer responsibility~\cite{makati2024Promise,hartman2025evaluating,overlayfactsheet}. Unlike such approaches, \systemname{} does not conceal these errors.}

\nan{Our findings highlight that designing accessible tools from the outset benefits both users and the effectiveness of assistive agents
~\cite{law2005programmer,perera2025sky}. Involving SR users in design and testing helps ensure that tools meet their needs~\cite{shinohara2018tenets,makati2024Promise}, while AI can further support this process~\cite{huq2023a11ydev}---by assisting in accessible design, generating code that follows accessibility standards~\cite{Mowar2025CodeA11y,mowar2024tab}, or detecting issues during testing~\cite{Taeb2024AXNav}. Achieving this vision will require models that are more aware of accessibility principles and SR interaction patterns. Ultimately, better accessible design and more capable agents can evolve together to create a more inclusive digital ecosystem.}

\subsection{Design Implications for Accessible Computing}

Our user study of \systemname{} demonstrates that context-aware guidance can significantly improve SR users' experiences and performance. 
Based on these findings, several design implications emerge for future accessible computing systems: 

\begin{itemize}[topsep=0pt,left=5pt]
    
    \item \textbf{Ensure context-awareness:} 
    \systemname{} shows that automatically collecting and integrating multiple sources of contextual information can greatly enhance the effectiveness of guidance for SR users. Context-aware systems should be able to understand users' current location in an interface, recent interactions, and relevant dynamic updates. 
    Future AI systems should capture a holistic set of contextual cues and intelligently manage the context relevant to each user query to build a shared understanding between the system and the user.

    \item \textbf{Generate accessible responses:} 
    AI-generated guidance should be designed with accessibility as a core principle. This means that the content itself---its wording, structure, and conveyed information—must be understandable, actionable, and free of barriers for users~\cite{huh2025vid2coach}. For example, referring to an interface element by a vague visual description such as ``black button'' may be confusing for SR users, whereas specifying a clearly labeled ``Confirm'' button is more accessible. Accessible responses ensure that users can interpret instructions, identify interface elements, and complete tasks without ambiguity, supporting equitable use of digital systems for all users.


    \item \textbf{Minimize disruption in interaction design:} 
    Participants valued seamless access to assistance without switching windows or losing focus. 
    The manner in which the assistant delivers its responses is as important as the content itself.
    Assistance should be invoked via keyboard shortcuts, presented in transient, non-modal dialogs, and delivered incrementally in individually navigable steps. 
    Systems should also allow users to signal confusion or request clarification with a single, simple action, without disrupting the overall guidance flow.

    \item \textbf{Provide transparency \fix{for building trust}:} 
    \nan{Our findings show that user} trust in AI assistants is built gradually through consistent, reliable responses\nan{, consistent with prior work~\cite{bansal2024challenges,nielsen2023ai}.
    Early skepticism and ongoing concerns about hallucinations or incorrect steps highlight that trust is provisional and can be fragile, especially for SR users who may struggle to verify information~\cite{chen2025screenreaderusersvibe,perera2025sky}. 
    To support trust, future systems should provide mechanisms that allow users to understand and verify AI behavior rather than treating it as a black box. 
    Potential solutions include offering accessible explanations of the AI's process (e.g., clear audio cues during processing)~\cite{chen2025screenreaderusersvibe}, providing brief rationales for suggestions~\cite{alharbi2024misfitting}, and cross-checking outputs across models~\cite{perera2025sky,adnin2024look}.}


\end{itemize}

\subsection{Limitations and Future work}

While our work shows promising results in supporting SR users' computer use, it has several limitations. 

\nan{First, our study was focused on participants who were not familiar with the target software, allowing us to observe how they seek assistance. However, this design limits our ability to understand how users who are already proficient with an application might leverage their prior knowledge; future work could investigate this through in-the-wild deployments. 
Additionally, the recruited SR users generally had moderate experience with LLMs tools, which was not a selection criterion but rather a characteristic of those who volunteered. As a result, the findings may be biased toward more technically confident users. While this group represents early adopters likely to engage with emerging assistive AI systems, future work should include SR users with little or no LLMs tools experience and employ longitudinal designs to examine how interactions, usage, and reliance on AI assistants evolve over time~\cite{chen2025screenreaderusersvibe}.}



\nan{Observations from unideal cases (\autoref{sec:unideal}) reveal several opportunities for improvement. Future systems could enhance environmental awareness, for example by detecting shortcut conflicts and cropping screenshots based on areas of interest to improve resolution. 
Such enhancements could reduce errors caused by incomplete or coarse-grained information. 
While \systemname{} currently relies on manually collected documentation to provide precise guidance for the software, this approach limits its coverage. Future work could explore automated collection methods and leverage richer data modalities to reduce model hallucinations~\cite{huh2025vid2coach}. 
Moreover, fine-tuning models to better understand SR usage could improve response quality, \final{while future work could explore compute-efficient local deployment to support more economical assistance and wider adoption~\cite{wu2025llm}.}
Achieving this would likely require more annotated data, careful fine-tuning strategies, and model architecture optimization~\cite{yang2024harnessing,pratap2025fine,pope2023efficiently}.
In addition, systems could proactively identify points where users are likely to become confused and provide timely guidance, while carefully avoiding unnecessary interruptions~\cite{flores2025impact}, complementing adaptive support.}

\nan{Our study also revealed that}
users sometimes struggled to articulate their goals, especially when interacting with unfamiliar tasks~\cite{perera2025sky}. 
Future systems could incorporate clarification mechanisms, \nan{
such as proactively asking follow-up questions~\cite{peng2025morae} or providing prompt scaffolding to guide users who are unfamiliar with task-specific terminology.} 
Moreover, the implementation is tied to NVDA on Windows. Extending \systemname{} to support multiple platforms, assistive technologies (e.g., JAWS, VoiceOver), and diverse user population with different accessibility needs would broaden its impact.

Finally, \nan{to enable context-aware support,} the system captures \nan{contexts (e.g., screenshots and SR traces)} to understand their interactions, raising privacy and ethical concerns. 
Future work should adopt privacy-by-design principles to mitigate these risks \fix{while maintaining usability.}
\nan{Essential approaches include using local models that run on the user's device to avoid transmitting sensitive information to external servers~\cite{team2025gemma}, or applying privacy-preserving techniques (e.g., encrypted or anonymized models) before sending data to remote servers for further processing~\cite{madhusudhanan2025privacy}.}

\section{Conclusion}
\label{sec-conclusion}
We presented \systemname{}, an on-demand, LLM-powered assistant that helps SR users navigate and interact with digital systems more effectively. 
By integrating multiple sources of context---live screen reader traces, screen states, and recent chat history---\systemname{} generates guidance that is precise, accessible, and aligned with SR user practices.
Our study with 12 participants showed that it not only improved task completion performance but also reduced effort and frustration. 
Further, \systemname{} demonstrates how AI-driven assistive technologies can lower barriers for users with visual impairments in computer use and envision more equitable participation in education, work, and everyday computing in the future.
These findings highlight the value of context-aware, screen reader-oriented design in developing future inclusive assistive systems.

\begin{acks}
\final{We thank the anonymous reviewers and the associate chair for their constructive feedback. Siming Chen acknowledges the following grant: Natural Science Foundation of China (NSFC No.62472099).}
\end{acks}


\bibliographystyle{ACM-Reference-Format}
\bibliography{main}


\end{document}